\documentclass[aps,physrev,twocolumn,groupedaddress]{revtex4-2}

\usepackage[switch]{lineno} 

\usepackage{graphicx}

\begin{document}


\title{Formation of dusty clumps in the torus of active galactic nuclei}



\author{Xinwu Cao}
 \homepage{https://person.zju.edu.cn/en/0019144}
 \email[]{xwcao@zju.edu.cn}
\affiliation{Institute for Astronomy, Center for Cosmology and Computational Astrophysics, School of Physics, Zhejiang University, 866 Yuhangtang Road, Hangzhou, 310058, People’s Republic of China}
\author{Renyue Cen}
\affiliation{Center for Cosmology and Computational Astrophysics, Institute for Astronomy, School of Physics, Zhejiang University, 866 Yuhangtang Road, Hangzhou, 310058, People’s Republic of China}
\author{Qingwen Wu}
\affiliation{Department of Astronomy, School of Physics, Huazhong University of Science and Technology, Luoyu Road 1037, Wuhan, People's Republic of China}
\author{Jiancheng Wu}
\affiliation{Institute for Astronomy, School of Physics, Zhejiang University, 866 Yuhangtang Road, Hangzhou, 310058, People’s Republic of China
}




\date{\today}

\begin{abstract}
The putative dusty torus is a key ingredient of the unification scheme of active galactic nuclei (AGN), but its origin remains a mystery. Here we put forward a new physical model to explain how a large number of small dusty gas clumps form and they collectively appear as a geometrically thick dynamic dusty torus. The circumnuclear hot gas flows towards the central black hole (BH) and forms a rotating   disk on sub-pc scales. A fraction of inflowing hot gas condenses to form small cold clumps due to thermal instabilities, when the accretion rate is sufficiently high. These cold dusty gas clumps are irradiated by the central accretion disk and re-radiate as dust emission mostly in the infrared. We propose that the dusty torus in AGN consists of such cold clumps vertically supported by the radiation force against gravity. For clumps with suitable column density, the vertical component of the BH gravity is in quasi-static equilibrium with the infrared radiation force together with the vertical component of the   disk radiation force. Our model is robust in the sense that for any reasonable range of parameters concerning clump vertical dynamical equilibrium a torus exists. We further show that the hot gas in the rotating flow condenses to cold clumps only if its accretion rate is higher than about one percent of the Eddington rate. The radiation force is unable to lift the cold gas clumps up away from the mid-plane when the luminosity of the   disk surrounding the BH is lower than 0.1 percent of the Eddington luminosity. These two features of our model may provide a physical explanation for the lack of evidence of dusty tori in low-luminosity AGNs.
\end{abstract}


\maketitle


\section{Introduction} \label{sec:intro}

Almost every galaxy has a massive black hole (BH) at its center, which is observed as an active galactic nucleus (AGN) if a large amount of gas is being accreted onto the BH. AGNs are classified as type 1 or type 2 according to their optical spectra, i.e., type 1 AGNs have broad emission lines, while no broad emission line appears in the spectra of type 2 AGNs. A dusty torus with toroidal structure is assumed in the unification scheme of AGNs, which blocks the broad line region (BLR) emssion of the type 2 AGNs if it is viewed at a large angle with respect to the axis of the rotating system, while those observed pole-on appear as type 1 AGNs \citep[][]{1985ApJ...297..621A,1993ARA&A..31..473A,1995PASP..107..803U}. There is abundant observational evidence of the dusty torus in AGNs \citep[see][for recent reviews, and the references therein]{2015ARA&A..53..365N,2017NatAs...1..679R}. The high resolution mid-infrared imagining and interferometry give constraints on such toroidal structure with an outer radius of $0.1-10$~pc in some nearby AGNs \citep[][]{2004Natur.429...47J,2005ApJ...618L..17P,2007A&A...474..837T,2008ApJ...681..141R,2022Natur.602..403G}.

The infrared radiation emitted by the dusty torus is a reprocessed fraction of the accretion   disk emission, so the information of the geometry, i.e., the solid angle subtended by the dusty torus, can be estimated with  
the ratio of the torus to the   disk luminosity, which indicates a geometrically thick dusty torus in AGNs 
\citep[][]{2005ApJ...619...86C,2013MNRAS.430.3445M,2013ApJ...773..176G,2013ApJ...777...86L,2016MNRAS.455.3968H,2016MNRAS.458.2288S,2018ApJ...862..118Z,2021ApJ...912...91T}. 
This is consistent with that estimated from the fraction of quasars in the Third Cambridge Revised Catalog of Radio Sources (3CR) sample \citep[][]{1996ApJ...462..163H}.

The dusty torus is irradiated by the   disk emission, and the dust cannot survive inside a critical radius at which the
temperature is higher than $\sim 1500$~K. The dust begins to sublimate, and the torus is truncated within this radius, which can be estimated as $R_{\rm torus}\simeq 0.06(L_{\rm   disk}/10^{38}~{\rm W})^{1/2}$~pc \citep[][]{1993ApJ...404L..51N,1996ApJ...462..163H}. High cadence monitoring observations on the Seyfert 1 galaxy NGC 4151 indicate a lag time of $\sim 48$ days between the $V$ and $K$ light curves, corresponding to an inner radius $\sim 0.04$~pc of the dusty torus, which is roughly consistent with the estimated evaporation radius of the dust \citep[][]{2004ApJ...600L..35M}.

For a geometrically thick torus with relative thickness $H/R\sim 1$ at the radius of sub-pc scale, the random velocity should be $\gtrsim 100~{\rm km~s}^{-1}$, if the vertical extension of the torus is maintained by the chaotic motion of the matter. The temperature is about $\sim 10^6$~K for this velocity, which is too high for dust to survive. \citet{1988ApJ...329..702K} suggested that the cold dust can be in clouds, which move at a random velocity  of $\gtrsim 100~{\rm km~s}^{-1}$, {though how such clouds form in the torus is still unclear in this scenario}. 
However, the high random velocity of the clouds in a geometrically thick torus leads to too many collisions between clouds on the assumption that the covering factor of the clouds is of order unity, which may feed too much gas onto the BH. Strong high energy $\gamma$-ray emission due to such frequent collisions between clouds has been predicted by the calculations in \citet{2004ApJ...614L..21W}, {which has been detected by Fermi-LAT \citep[][]{2022ApJS..260...53A}. However, it was suggested to be associated with knots within the jet-like structure of NGC 1068 \citep[][]{2024A&A...687A.139S}.} In order to solve this issue, an alternative model was suggested, in which a smooth torus is puffed up by the infrared radiation of the gas irradiated by the   disk emission \citep*[][]{1992ApJ...399L..23P,2007ApJ...661...52K,2017ApJ...843...58C}, whereas a model of discrete clumps is required to reproduce the infrared SEDs, and some evidence of inhomogeneities of gas has been found in the infrared and X-ray observations \citep[][]{2009ApJ...702.1127R,2023MNRAS.522.4098K}. 
{The radiation hydrodynamic simulations of such a smooth dusty gas flow show that a geometrically thin   disk forms instead of a thick torus, which is almost supported by thermal pressure \citep[][]{2016MNRAS.460..980N}. Contrary to the model of \citet{2007ApJ...661...52K}, the radiation pressure by infrared photons is not sufficient to puff the dusty   disk vertically.} However, it is worthy to point out that the concept and formulation given in these works of smooth torus are also valid for clumpy dusty torus provided appropriate opacity is adopted. Similar mechanism has also been proposed to play an important role in the formation of broad-line regions (BLRs) in AGNs \citep[][]{2011A&A...525L...8C,2017ApJ...846..154C,2018MNRAS.474.1970B}.

The origin of such clumpy dusty tori in AGNs is still a mystery, though it is found that a torus-like geometry of cold gas is formed in the numerical simulations on the interaction of AGN feedback and the interstellar medium (ISM) \citep{2016MNRAS.458..816H,2025OJAp....8E..56H}. It was suggested that the structure of the obscuration is more complex than a donut, instead, a model of an equatorial dense region accompanied by polar dusty outflows has been developed based on multi-waveband observations of AGNs \citep[][]{2012ApJ...755..149H,2013ApJ...771...87H,2014A&A...563A..82T,2018ApJ...862...17L,2020ApJ...900..174V}.    
The basic driving mechanism for launching such outflows is radiation pressure on dust, as the dust opacity is much higher than the Thomson opacity \citep[e.g.,][]{2012ApJ...755..149H}. The trajectories of dusty gas clumps have been calculated for radiatively accelerated dusty outflows, in which both gravity and the AGN radiation as well as the re-radiation by the hot, dusty gas clouds themselves are properly included \citep[][]{2012MNRAS.424..820B,2019ApJ...884..171H,2020ApJ...900..174V}. The resulting morphology consisting of a disk of material and a hyperboloid polar wind launched from the inner edge of the dusty disk is consistent with high-angular resolution infrared and sub-mm observations of some local Seyfert AGNs. Radiation hydrodynamical simulations for the obscuring structure in AGNs have been carried out by some workers \citep[][]{2011ApJ...741...29D,2012ApJ...747....8D,2012ApJ...758...66W,2015ApJ...812...82W,2019ApJ...876..137W,2020ApJ...897...26W}, in which geometrically thick structure and inhomogeneities are indeed produced.  

In this paper, we propose a model of the formation of the dusty torus in the hot gas flow. {In this model, the cold gas is condensed from the hot gas into clumps due to thermal instabilities. Unlike the infrared radiation supported smooth torus model of \citet{2007ApJ...661...52K}, our model shows that the cold gas clumps with suitable column density can be vertically supported by the   disk radiation force together with infrared radiation force of the torus irradiated by the   disk, which can be in vertical quasi-static state. The trajectories of individual clumps are vertically layered. It therefore avoids frequent collisions between clouds in \citet{1988ApJ...329..702K}'s torus model.} The model and the results are given in Section \ref{sec:model}. Section \ref{sec:discussion} contains the discussion of the model.

\section{Model}\label{sec:model}

The circumnuclear hot gas in the galaxies is influenced by the gravity of the central massive BH, and therefore it falls onto the BH, which can be described approximately by the Bondi accretion if the angular momentum of the gas is sufficiently low \citep*[][]{1952MNRAS.112..195B}. The hot gas finally forms as a rotating flow in the region with radius much smaller than the Bondi radius, as the gas at the Bondi radius may inevitably possess a certain amount of angular momentum, though it may be substantially lower than the Keplerian value, The hot gas in this rotating flow may suffer from the non-linear thermal instability, which makes the hot gas condense to cold gas clumps, if the density of the hot gas is high, and its cooling timescale is short \citep*[][]{2012ApJ...746...94G,2012MNRAS.419.3319M,2012MNRAS.420.3174S,2013MNRAS.432.3401G}. {The dust grains are formed in the cold gas clumps with temperature $\lesssim 1500~{\rm K}$ through coagulation and accretion of metal atoms and molecules. It is found that the dust formation timescale is much shorter than the free-fall timescale in the cold clumps, and therefore the formation of dust grains in the cold clumps is very efficient (see Section \ref{sec:condense}).} The cold dusty clumps decouple from the hot gas, and they may co-exist and rotate with the BH. 

 {It has been suggested that dust plays an important role in launching outflows by radiation pressure in quasars \citep[see][and the references therein]{2024MNRAS.533.4384I}, due to its high opacity.}  As indicated by the observations, geometrically thick dusty tori consisting of dusty clumps are  present in AGNs. One key issue in the torus scenario is how the torus can be maintained as a geometrically thick shape. As discussed in the Introduction, a geometrically thick torus maintained by chaotic motion of clumps is quite unlikely. 

{Here, we first analyze the possibility of the dusty clumps being supported by the radiation force of the accretion disk. Suppose a clump locating at a distance of $R$ from the BH, the gravity and the radiation force per unit area of the clump are 
\begin{equation}
   F_{\rm g}={\frac {GM_{\rm bh}\Sigma_{\rm cl}}{R^2}},\label{force_g_z}
\end{equation}
and
\begin{equation}
   F_{{\rm rad}}={\frac {L_{\rm   disk}(1-e^{-\tau_{\rm cl}^{\rm uv}})}{4\pi R^2 c}}, \label{force_rad_z}
\end{equation}
respectively, where $\Sigma_{\rm cl}$ is the surface density of the clump, and its optical depth $\tau_{\rm cl}^{\rm uv}=\Sigma_{\rm cl}\kappa_{\rm uv}$. {The opacity $\kappa_{\rm uv}=1200~ {\rm cm^2~g}^{-1}$ for the UV photons from the disk with a peak flux around $1000~{\rm \AA}$ \citep[see Figure 9 in][]{2003ARA&A..41..241D}, which is an updated version of \citet{2001ApJ...554..778L}.} In the case of the clump in force equilibrium, i.e., $F_{\rm g}=F_{\rm rad}$, we have 
\begin{equation}
    {\frac {GM_{\rm bh}\Sigma_{\rm cl}}{R^2}}={\frac {L_{\rm   disk}(1-e^{-\tau_{\rm cl}^{\rm uv}})}{4\pi R^2 c}}. \label{force_balance_0}
\end{equation}
There is a maximal vertical radiation force when $\tau_{\rm cl}^{\rm uv}\rightarrow \infty$ (see Equation \ref{force_rad_z}), while the gravity always increases with the surface density $\Sigma_{\rm cl}$, which means that the balance between the gravity and radiation force cannot be maintained if $\Sigma_{\rm cl}$ is too large. Thus, the clump with surface density higher than a critical value $\Sigma_{\rm cl}^{\rm max}$ will inevitably moves inwards for a given incident radiation flux from the disk, which is given by
\begin{equation}
    {\frac {GM_{\rm bh}\Sigma_{\rm cl}^{\rm max}}{R^2}}={\frac {L_{\rm   disk}}{4\pi R^2 c}}, \label{force_balance}
\end{equation}
for an optically thick clump ($\tau_{\rm cl}^{\rm uv}\gg 1$). With this equation, we derive a relation of the maximal  column density of the clump $N_{\rm cl}^{\rm max}$ with the Eddingtion ratio of the disk as
\begin{equation}
    N_{\rm cl}^{\rm max}={\frac {\lambda_{\rm d}}{\mu_{\rm cl}m_{\rm p}\kappa_{\rm T}}}=1.17\times 10^{24}\lambda_{\rm d}~{\rm cm}^{-2}, \label{n_cl_0}
\end{equation}
in which $\lambda_{\rm d}=L_{\rm   disk}/L_{\rm Edd}$, and $L_{\rm Edd}=4\pi GM_{\rm bh}c/\kappa_{\rm T}$ are used. The column density $N_{\rm cl}=(\kappa_{\rm uv}\mu_{\rm cl}m_{\rm p})^{-1}=3.89\times 10^{20}~{\rm cm}^{-2}$, or $\lambda_{\rm d}\approx 5.3\times 10^{-4}$, corresponding to $\tau_{\rm cl}^{\rm uv}=1$. Thus, an optically thick clump with $N_{\rm cl}\sim 10^{21}-10^{24}~{\rm cm}^{-2}$ can be sustained by the radiation force, provided the disk luminosity is $\sim 10^{-3}-1$ Eddington luminosity.  }

{If the column density of the clump is sufficiently low, it becomes optically thin ($\tau_{\rm cl}^{\rm uv}\ll 1$). Equation (\ref{force_balance_0}) can be reduced to 
\begin{equation}
    {\frac {GM_{\rm bh}\Sigma_{\rm cl}}{R^2}}={\frac {L_{\rm   disk}\tau_{\rm cl}^{\rm uv}}{4\pi R^2 c}}={\frac {L_{\rm   disk}\Sigma_{\rm cl}\kappa_{\rm uv}}
    {4\pi R^2 c}}, \label{force_balance_2}
\end{equation}
which leads to $\lambda_{\rm d,min}=\kappa_{\rm T}/\kappa_{uv}=3.3\times 10^{-4}$. It indicates that the radiation force can never counter the gravity for a clump with arbitrary column density if $\lambda_{\rm d}<\lambda_{\rm d,min}$ (the right term is always less than the left one in Equation \ref{force_balance_2}). Our simple analysis shows that the cold clumps with suitable column density may be sustained by the radiation force in quasi-equilibrium state, as long as the Eddington ratio of the cold disk surrounding the BH is sufficient large. The detailed calculations for rotating gas clumps including the infrared radiation force due to the dusty torus will be given in Section \ref{sec:rad_torus}.} 

\begin{figure}
	\centering
	\includegraphics[width=0.9 \columnwidth]{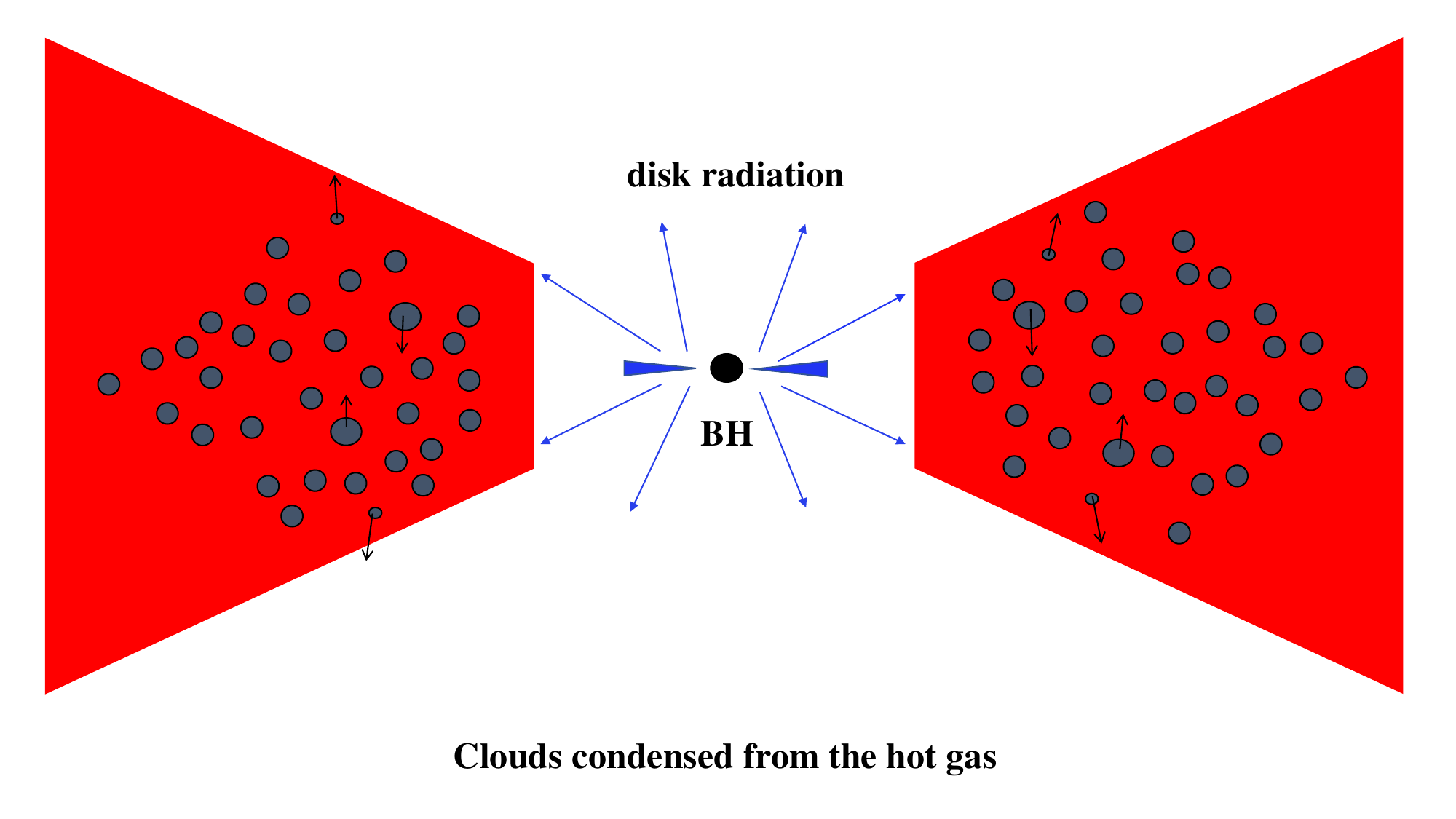}
	\caption{Illustration of the model (not to scale in size and number of the clumps). Only the clumps with suitable column density can be vertically supported by the radiation force against the gravity, while the remainder are either driven away or sinking down. Some clumps may fall onto the BH through a cold   disk. }
	\label{fig:illustr}
\end{figure}

For the clumps condensed from the hot gas in a rotating flow, the clumps are inevitably rotating, which are irradiated by the disk emission. They re-radiate in infrared wavebands, and thus the clumps with suitable column density may be vertically supported by the disk radiation together with the reprocessed infrared emission, which leads to a geometrically thick spacial distribution of the clumps. These rotating clumps are in vertically quasi-static equilibrium {in the sense that the relative velocity of the clouds is smaller than the sound speed of the clouds}, which avoid frequent collisions between clumps. We suggest that these clumps may block a large fraction of the disk radiation, which emerge as the dusty torus of AGN inferred from the observations (see Figure \ref{fig:illustr} for the illustration of the model).     

{The hot gas co-existing with cold clumps will finally inflows into the inner region near the BH, which may appear as an advection dominated accretion flow (ADAF) if its accretion rate is low, otherwise, the gas may cool down to form a cold thin disk surrounding the BH \citep[][]{1995ApJ...452..710N}. In the case of a cold disk surrounding the BH, the accretion rate of the cold disk should be higher than that of the hot gas flow in the outer region for the steady case (or averaged over a long period of time), as some additional ambient cold gas may also be fed onto the cold disk.}

\subsection{Condensation of the hot gas in the accretion flow}\label{sec:condense}

There is plenty of observational evidence of the extended hot gas in the circumnuclear region of the galaxies, which is supposed to be the reservoir to feed the central massive BHs in AGNs, e.g., radio galaxies and Seyfert galaxies \citep[][]{2006MNRAS.372...21A,2010ApJ...719L.208W,2011ApJ...729...75W,2022ApJ...938..127X}, or even in normal galaxies, including our own galaxy \citep[e.g.,][]{2004ApJ...613..322Q}.  

The Bondi accretion radius of the circumnuclear hot gas is
\begin{equation}
    R_{\rm B}={\frac {2GM_{\rm bh}}{c_{\rm s,B}^2}}, \label{r_b}
\end{equation}
where the sound speed $c_{\rm s,B}=(\gamma k_{\rm B}T_{\rm B}/\mu m_{\rm p})^{1/2}$, $T_{\rm B}$ is the gas temperature at the Bondi radius, and $\gamma=5/3$ is adopted for adiabatic accretion \citep*[][]{1952MNRAS.112..195B}. 

{Diffuse X-ray emission detected in some Seyfert galaxies indicates circumnuclear hot gas in $\sim {\rm kpc}$ scale \citep[e.g.,][]{2010ApJ...719L.208W,2011ApJ...729...75W,2022ApJ...938..127X}. In the Seyfert galaxy NGC~4151, diffuse soft X-ray emission extending from the active nucleus has been observed. The background subtracted surface brightness profile can be fitted with one-dimensional $\beta$-model, the best-fit index of which gives $\beta=0.39$ \citep[see Figure 3 in][]{2010ApJ...719L.208W}. Thus, the density of the hot gas is $\propto R^{-3\beta}\sim R^{-1.17}$ \citep[][]{1986ApJ...310..637T}. {As the emissivity is proportional to the square of density, the extended X-ray emission of this source is predominantly from the outer region.  The luminosity of the extended soft X-ray emission, $L_{\rm 0.5-2~keV}\sim 10^{39}~{\rm erg~s}^{-1}$ \citep{2010ApJ...719L.208W}, can be used to estimate the electron number density of the hot gas at 1~kpc with $L_{\rm 0.5-2~keV}\sim 4\pi R^3\epsilon$ ($R=1~{\rm kpc}$), where the emissivity $\epsilon\approx 6.2\times 10^{-19}T^{-0.6}n_{\rm e}^2~{\rm erg~s^{-1}~cm^{-3}}$ for the thermal gas with temperature $kT=0.25$~keV \citep{1977ApJ...215..213M}. The inferred electron number density $n_{\rm e}\approx 5.7\times 10^{-3}~{\rm cm}^{-3}$ at $1~{\rm kpc}$.} 

Adopting electron number density estimated with the X-ray observations, the density at the Bondi radius $R_{\rm B}$ can be estimated as $n_{\rm e}(R_{\rm B})=n_{\rm e}(R=1~{\rm kpc})(R_{\rm B}/1~{\rm kpc})^{-1.17}$. The central BH mass of NGC~4151, $M_{\rm bh}\simeq 3\times 10^7~M_\odot$ \citep[][]{2021ApJ...916...25R}. Assuming the gas temperature $kT_{\rm B}=1~{\rm keV}$ at the Bondi radius, one can estimate the Bondi radius with Equation (\ref{r_b}) as $R_{\rm B}\approx 1.61~{\rm pc}$. The Bondi accretion rate is  }
{\begin{equation}
    \dot{M}_{\rm B}=4\pi \lambda (GM_{\rm bh})^2\rho_{\rm B}c_{\rm s,B}^{-3}, \label{mdot_bondi}
\end{equation}}
{where $\lambda=0.25$, and $\rho_{\rm B}$ is the density of the gas at the Bondi radius $R_{\rm B}$ \citep[][]{2006MNRAS.372...21A}. {The accretion rate of NGC~4151 is therefore estimated as $\sim 8.6\times 10^{21}~{\rm g~s}^{-1}$ with the density derived from the X-ray observation. Assuming a conventional radiation efficiency $\eta_{\rm rad}=0.1$, its accretion luminosity is $\sim 7.7\times 10^{41}~{\rm erg~s}^{-1}$, which is significantly lower than the observed bolometric luminosity of NGC~4151, $L_{\rm bol}\sim 7.3\times 10^{43}~{\rm erg~s}^{-1}$ \citep[][]{2005ApJ...629...61K}. It seems that the BH in NGC 4151 is fed mostly by the gas much colder than the X-ray emitting hot gas.} 

{We note that the measured slope of the radial distribution of hot gas in this source is very shallow. If its slope is steep as those observed in some other galaxies with $\beta>0.5$ \citep{1986ApJ...310..637T}, the density of the gas at the Bondi radius will be substantially increased. The Bondi accretion rate becomes $\sim 9.8\times 10^{23}~{\rm g~s}^{-1}$, and accordingly its accretion luminosity is $\sim 8.9\times 10^{43}~{\rm erg~s}^{-1}$, for a higher $\beta=0.6$, if the electron density remains unchanged, which is comparable with the observed bolometric luminosity of NGC~4151. We conjecture that such situation may be the case in some other AGNs.}

The inflowing hot gas can be described fairly well by the Bondi accretion model, if its angular momentum is sufficiently low. The high resolution X-ray observations on a sample of radio galaxies show that the typical temperature of the gas $\sim$keV at $R_{\rm B}$, and the Bondi radii are in the range of $\sim 1-100$pc \citep[][]{2006MNRAS.372...21A}, which are significantly larger than the size of the torus inferred from the observations of AGNs. The gas rotating with a small angular velocity at the Bondi radius falls almost freely to the BH until the circularization radius, which may be located outside the dusty torus, if the angular velocity of the gas is $\gtrsim (R_{\rm torus}/R_{\rm B})^{1/2}$ times of the Keplerian velocity at the Bondi radius. In this case, a rotating hot gas  disk may incubate cold clumps in the region of $R\gtrsim R_{\rm torus}$. 

{Some recent hydrodynamic simulations suggest that the accretion flow with a small angular momentum results in a cold   disk within the circularization radius due to the thermal instabilities of the hot gas, as long as the mass accretion rate is higher than $\sim 10^{-3}$ Eddington rate \citep[][]{2019MNRAS.486.5377I,2022ApJ...926...50T}. We notice that the viscosity parameter $\alpha\sim 0.01$ is adopted in their simulations. The density of the accretion flow is inversely proportional to the value of $\alpha$, and the cooling of the gas is approximately proportional to the square of density. Thus, the critical accretion rate for an hot accretion flow is roughly proportional to $\alpha^2$, which is similar to the transition accretion rate of an ADAF to a thin disk \citep[][]{1995ApJ...452..710N}. There is plenty of observational evidence that such accretion mode transitions usually occur at $\sim 0.01$ Eddington luminosity in AGNs, which corresponds to $\alpha\sim 0.1$ expected by the ADAF model \citep[see][for a review]{2014ARA&A..52..529Y}.  In this work, we adopt $\alpha=0.1$, which allows a hot flow accreting at a rate much higher than $\sim 0.01$ Eddington rate in the outer region far away from the BH, as the cooling timescale of the gas in the outer region is much longer than that near the BH.}

The radial velocity of a rotating hot gas   disk is 
\begin{equation}
    v_{R,\rm h}=-\alpha c_{\rm s,h}{\frac {H_{\rm h}}{R}}, \label{v_r_h}
\end{equation}
where $H_{\rm h}$ is the half-thickness of the  disk, the isothermal sound speed $c_{\rm s,h}=H_{\rm h}\Omega_{\rm K}$, and the $\alpha$-viscosity is adopted. The temperature and the density of the hot gas   disk are
\begin{equation}
    T_{\rm h}={\frac {\mu m_{\rm p}}{k_{\rm B}}} \left({\frac {H_{\rm h}}{R}}\right)^2 v_{\rm K}^2=1.12\times 10^{13}r^{-1}\tilde{H}_{\rm h}^{2}~{\rm K}, \label{temp_h}
\end{equation}
and
\begin{displaymath}
  \rho_{\rm h}=-{\frac {\dot{M}_{\rm h}}{4\pi R H_{\rm h}v_{R,\rm h}}}  ~~~~~~~~~~~~~~~~~~~~~~~~~~~~~~~~~~~~~~~~~~~~~~~
\end{displaymath}
\begin{equation}
    =1.70\times 10^{-4}\alpha^{-1}m^{-1}\dot{m}_{\rm h}r^{-3/2}\tilde{H}_{\rm h}^{-3}~ {\rm g~cm}^{-3}, \label{rho_h}
\end{equation}
respectively, where $\tilde{H}_{\rm h}=H_{\rm h}/R$, $\dot{m}_{\rm h}=\dot{M}_{\rm h}/\dot{M}_{\rm Edd}$, $\dot{M}_{\rm h}$ is the accretion rate of the hot disk, and $\dot{M}_{\rm Edd}=L_{\rm Edd}/0.1c^2$. The pressure of the hot gas is
\begin{equation}
    p_{\rm h}=1.53\times 10^{17}\alpha^{-1}m^{-1}\dot{m}_{\rm h}r^{-5/2}\tilde{H}_{\rm h}^{-1} ~{\rm g~cm^{-1}~s^{-2}}. \label{p_h}
\end{equation}

It was shown that the non-linear growth of thermal instabilities may lead to condensation of cold clumps when $t_{\rm cool}/t_{\rm ff}\lesssim\xi_{\rm cf}$, where $t_{\rm cool}$ is the cooling timescale of the hot gas, $t_{\rm ff}$ is the free-fall timescale, and $\xi_{\rm cf}=10$ is suggested by the numerical simulations \citep[][]{2012ApJ...746...94G,2012MNRAS.419.3319M,2012MNRAS.420.3174S}. Although these simulations were carried out for the hot gas in a scale of sub-parsec to several kpc, the micro-physics of the hot gas is similar to the hot gas flow considered in this work in a smaller scale, and therefore we believe that the main conclusion derived from their simulations should still be valid for our present investigation on the hot accretion flow within the Bondi radius. In fact, previous works on the accretion flow, either analytic or numerical simulations, indeed show that an initially one-phase hot accretion flow rather quickly transitions to a two-phase/hot-cold acrretion flow due to thermal instabilities \citep[][]{1983ApJ...267...18K,2013ApJ...767..156M}.  

The cooling timescale is
\begin{equation}
t_{\rm cool}={\frac u {f_{\rm rad}^-}}, \label{t_cool}
\end{equation}
where the internal energy density $u=1.5\rho_{\rm h}k_{\rm B}T_{\rm h}/\mu m_{\rm p}$, and the emissivity {$f_{\rm rad}^-(\rho_{\rm h}, T_{\rm h})$ is calculated with the cooling function given by \citet{1993ApJS...88..253S} for solar abundance}. The free-fall timescale is \citep[][]{2013MNRAS.432.3401G}
\begin{equation}
    t_{\rm ff}=\left({\frac {2R^3}{GM_{\rm bh}}}\right)^{1/2}. \label{t_ff}
\end{equation}

{The temperature and density of the hot disk are derived with Equations (\ref{temp_h}) and (\ref{rho_h}), when the values of the hot disk parameters are specified, with which the cooling and free-fall timescales are calculated with Equations (\ref{t_cool}) and (\ref{t_ff}). The criterion for thermal instabilities $t_{\rm cool}/t_{\rm ff}\lesssim\xi_{\rm cf}$ sets a lower limit on the accretion rate of hot gas.} The growth of the thermal instability is highly non-linear. The cooling timescale of the cloud is $\propto 1/\rho^{2}$, while the internal energy $u\propto T$, and the density $\rho\propto 1/T$ as the cloud is always in pressure equilibrium with the ambient hot gas, which implies the real cooling timescale should be substantially lower than that estimated by using the hot gas temperature by Equation (\ref{t_cool}). This has been verified by the numerical simulations \citep[][]{2012MNRAS.419.3319M,2012MNRAS.420.3174S}. For simplicity, we adopt $\xi_{\rm cf}=10$ throughout this work, as suggested by the numerical simulations \citep[][]{2012MNRAS.419.3319M,2012MNRAS.420.3174S}. 

The cold clumps formed in the hot gas flow are irradiated by the disk radiation. The temperature of the clumps will be high if they are too close to the BH. The inner radius of the dusty torus is determined by the sublimation of the dust in the torus, which gives \citep[][]{1993ApJ...404L..51N,1996ApJ...462..163H}
\begin{displaymath}
R_{\rm torus}\sim 0.06
\left(
{\frac {L_{\rm   disk}}{10^{45}{\rm erg s^{-1}}}}
\right)^{1/2}~{\rm pc}~~~~~~~~~~~~~~~~~~~~~~~~~
\end{displaymath}
\begin{equation}
=2.12\times 10^{-2}\left({\frac {\lambda_{\rm d}}{0.01}}\right)^{1/2}\left({\frac m {10^8}}\right)^{1/2}~{\rm pc},\label{r_in}
\end{equation}
or
\begin{equation}
r_{\rm torus}={\frac {R_{\rm torus}}{R_{\rm g}}}\sim 4.43\times 10^3  \left({\frac {\lambda_{\rm d}}{0.01}}\right)^{1/2}\left({\frac m {10^8}}\right)^{-1/2},\label{r_in/r_g}
\end{equation}
where $L_{\rm   disk}$ is the   disk luminosity, $R_{\rm g}=GM_{\rm bh}/c^2$, and $m=M_{\rm bh}/M_\odot$, which is roughly consistent with the size  measured by the thermal dust reverberation method \citep[][]{2004ApJ...600L..35M}. Thus, the cold dusty clumps can be formed in the hot gas flow due to thermal instability only in the region of $R\ge R_{\rm torus}$.

The hot gas will finally feed the BH. Thus, {for the steady case, (or averaged over a long period of time)}, the mass accretion rate of the cold accretion disk surrounding the BH must be larger than the rate of hot rotating disk co-existing with the torus. i.e., 
\begin{equation}
   \dot{m}_{\rm d}=\lambda_{\rm d}\ge \dot{m}_{\rm h}(r_{\rm torus}), \label{mdot_cr_2} 
\end{equation}
if a typical radiation efficiency $\eta_{\rm rad}=0.1$ is assumed for the cold disk, because the circumnuclear cold gas may also contribute to the mass accretion in the cold disk surrounding the BH. We plot the relation of the minimal  accretion rate for thermal instabilities of the hot disk with the Eddington ratio of the cold disk at $R=R_{\rm torus}$ in Figure \ref{fig:arate_h_crit}. The hot gas will inevitably condense to cold clumps whenever the dimensionless accretion rate is greater than the critical rate. {The dotted line in Figure \ref{fig:arate_h_crit} indicates $\dot{m}_{\rm h}=\lambda_{\rm d}$. Therefore, the cold dusty clumps can be formed due to thermal instability only when the Eddington ratio $\lambda_{\rm d}$ of the cold disk is greater than a critical value (in the right side of the dotted line $\dot{m}_{\rm h}=\lambda_{\rm d}$), which may even higher if $\lambda_{\rm d}>\dot{m}_{\rm h}$. It is found that the cold gas clumps can be formed only if the the disk luminosity $\lambda_{\rm d}\gtrsim 3\times 10^{-3}-0.01$ (or higher in the case of $\lambda_{\rm d}>\dot{m}_{\rm h}$), for the BH mass in the range of $10^7-10^9~M_\odot$.  } 

\begin{figure}
	\centering
	\includegraphics[width=0.9 \columnwidth]{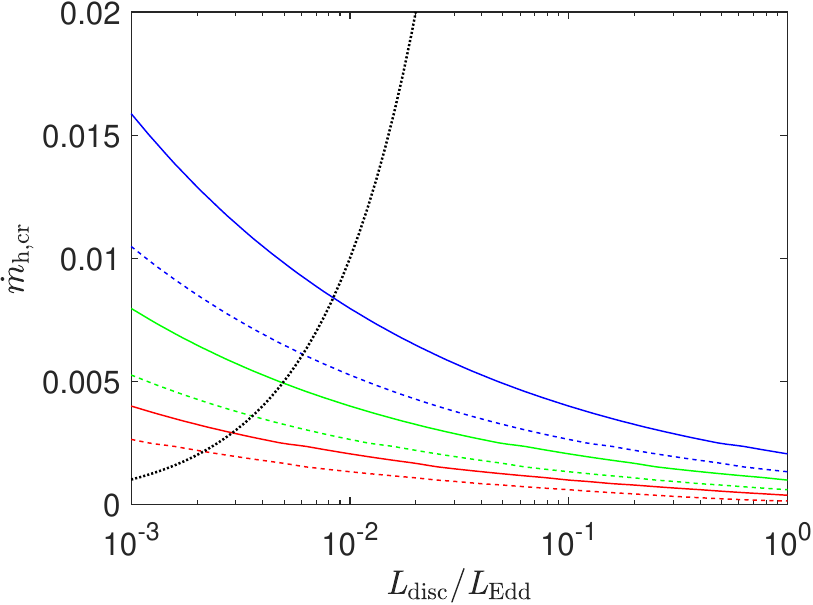}
	\caption{The critical accretion rate of the hot rotating   disk varies with the Eddington ratio of the cold   disk emission. The parameters, $\alpha=0.1$, and $\tilde{H}_{\rm h}=0.5$, are adopted in the calculations. The solid lines are the results calculated with $R=R_{\rm torus}$, while the dashed lines are for the cases of $R=2R_{\rm torus}$. The colored lines correspond to the results with different values of black hole mass, $m=10^7$ (red), $10^8$ (green), and $10^9$ (blue), respectively. The dotted line indicates $\dot{m}_{\rm h}=\lambda_{\rm d}$.  }
	\label{fig:arate_h_crit}
\end{figure}

\begin{figure}
	\centering
	\includegraphics[width=0.9 \columnwidth]{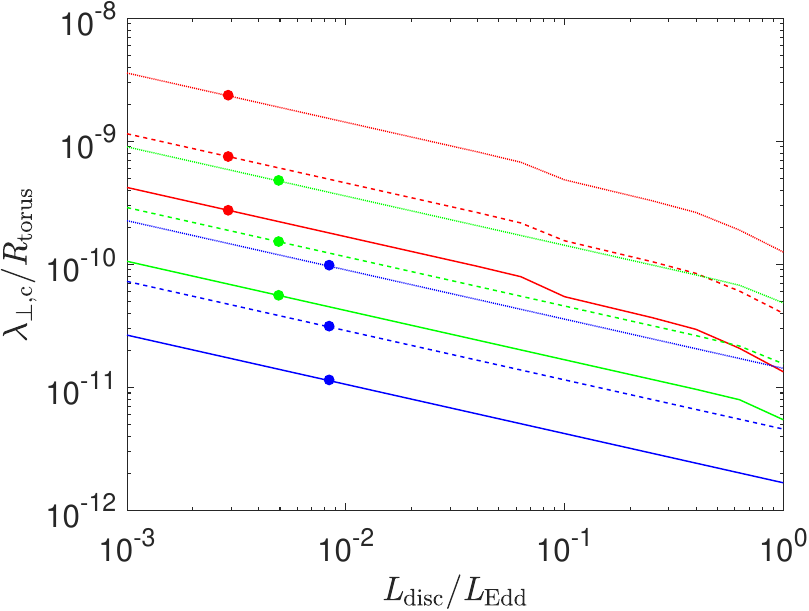}
	\caption{The critical wavelengths of perturbations normal to the magnetic field vary with the Eddington ratio of the cold disk emission. The thermal instability is suppressed by the thermal conduction only if the wavelength of the perturbation $\lambda<\lambda_{\perp,\rm cr}$. The parameters, $\alpha=0.1$, and $\tilde{H}_{\rm h}=0.5$, are adopted in the calculations. The solid lines are the results calculated with $\beta_{\rm h}=10$, while the dashed and dotted lines are for the cases of $\beta_{\rm h}=100$ and 1000 respectively. The dots indicate the critical Eddington ratios of the   disk luminosity.   
    The colored lines correspond to the results with different values of black hole mass, $m=10^7$ (red), $10^8$ (green), and $10^9$ (blue), respectively. }
	\label{fig:wavelenth_crit}
\end{figure}

\begin{figure}
	\centering
	\includegraphics[width=0.9 \columnwidth]{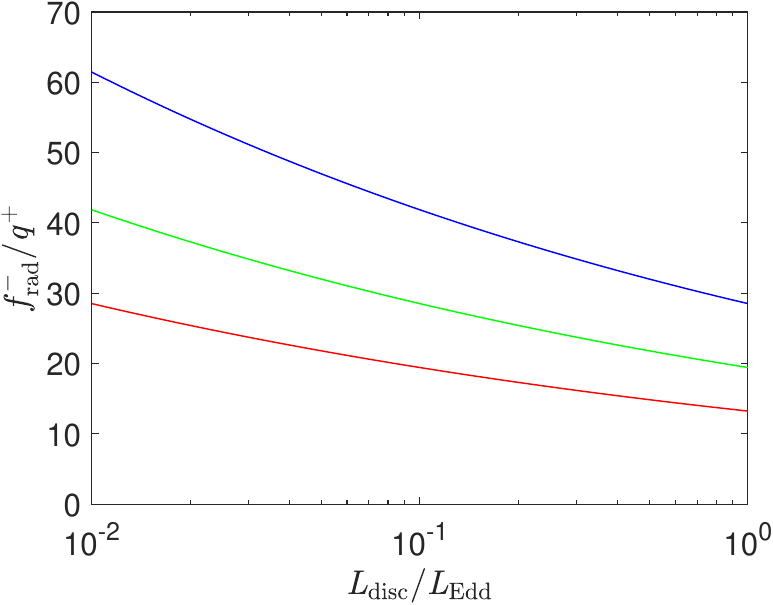}
	\caption{The ratios of the cooling rate to the heating rate of the gas with temperature $T=10^4$~K due to the incident radiation flux of the disk vary with $\lambda_{\rm d}$. The colored lines correspond to the results with different values of black hole mass, $m=10^7$ (red), $10^8$ (green), and $10^9$ (blue), respectively. }
	\label{fig:lambda_cool}
\end{figure}

{The thermal instability can be alleviated or even suppressed by the thermal conduction if the perturbation has an extremely short wavelength. The linear analysis on the thermal instability of the gas with magnetic fields is given in Appendix \ref{analysis_thermal_inst}.} We plot the critical wavelengths varying with the Eddington ratio of the cold   disk in Figure \ref{fig:wavelenth_crit}, where $\lambda_{\rm d}=\dot{m}_{\rm h}$ and $R=R_{\rm torus}$ are adopted.  The thermal instability is suppressed by the thermal conduction only if the perturbation has an extremely short wavelength. The thermal conduction is strongly anisotropic with respect to the magnetic field, and only short wavelength modes parallel to the magnetic field may be stabilized by the thermal conduction. The numerical simulations of galaxy formation have shown that, even very strong thermal conduction cannot suppress the thermal instability in the presence of magnetic field, and the growth of thermal instability via perturbations roughly orthogonal to the magnetic field is able to produce multi-phase gas structure. The criterion for thermal instability to produce multi-phase structure in the galaxy is therefore almost independent of the thermal conductivity \citep[see][for the detailed discussion]{2012MNRAS.419.3319M}. As the magnetic field of an accretion flow is usually strong, the plasma $\beta$ can be at the order of $\sim\alpha^{-1}$ \citep[][]{1991ApJ...376..214B}, which is much stronger than that of the hot gas in the galaxy far from the nucleus. Thus, we believe that two-phase structure is able to form through the thermal instabilities in the hot gas flow within the Bondi radius. 

 {One may worry that the cooling clouds would be stabilized at a temperature around $10^4$~K, as the clouds are irradiated by the   disk emission. This is indeed the case for the clouds in the broad-line regions (BLR) of AGNs \citep[][]{1990agn..conf...57N,2000ARA&A..38..521S}. However, the size of the torus is around an order of magnitude larger than the BLR size \citep[][]{2000ApJ...533..631K}, which means the incident flux irradiates the clouds in the torus is roughly two orders of magnitude weaker than that for BLR clouds. For the clouds in the torus, the heating rate due to the   disk emission is
\begin{equation}
q^+\sim {\frac {L_{\rm   disk}\rho_{\rm cl}(T)\kappa(T)}{4\pi R_{\rm torus}^2}}, 
\end{equation}
where the density of the cloud is calculated assuming it to be in pressure equilibrium with the hot gas if the temperature $T$ is specified. The opacity of the gas is a function of density and temperature given by \citet{2014MNRAS.445..479C}. With specified density and temperature of the gas, the cooling function is calculated with CLOUDY \citep{2023RMxAA..59..327C}. In Figure \ref{fig:lambda_cool}, we compare the heating rate due to the   disk emission with the cooling rate of the clouds in the torus assuming the gas temperature to be $10^4$~K. It is indeed found that the cooling rate always dominates over the heating rate due to the irradiation of the   disk emission, which means that the cloud will be continuously cooling down from $10^4$~K. }

 {The dust is formed in the clouds when the gas temperature is sufficiently low and the density is large enough, so-called the dust formation window. The one of the most efficient sources of dust is the winds of late-type giant and supergiant stars \citep[see][for a review, and the references therein]{2024A&ARv..32....2S}. The pressure of the inner edge of the cool shell is $\sim 10^{-4}~{\rm g~cm^{-1}~s^{-2}}$, and the pressure could be raised up to $\sim 10^{-2}~{\rm g~cm^{-1}~s^{-2}}$ by the shock waves of long-period variables \citep[][]{1999IAUS..191..159H}, which greatly increases the dust grain production \citep{1991A&A...242L...1F,1992A&A...266..321F}. \citet{2002ApJ...567L.107E} argued that the temperature and density of the gas in the outflows from AGNs are in the dust formation window established with the winds from giant stars \citep[][]{1991A&A...242L...1F,1992A&A...266..321F}, and they concluded that dust can be efficiently formed in the cooling outflows in AGNs. As the clouds in the torus are in pressure equilibrium with the ambient hot gas, their pressure can be estimated with Equation (\ref{p_h}), which is in the range of $\sim 10^{-2}-1~{\rm g~cm^{-1}~s^{-2}}$, substantially higher than that of the stellar winds.}

{We further estimate the growth of dust grains through accretion, which is described by 
\begin{equation}
    {\frac {da}{dt}}={\frac {w_a\rho_a\xi_a}{4\rho_s}}, \label{dadt}
\end{equation}
where the density of the dust grain $\rho_s\approx 1~{\rm g~cm}^{-3}$, $a$ is the radius of the dust grain, $w_a$ is the average velocity of the metal elements, and $\rho_a$ is the total density of all metal elements \citep[see][for the details]{1978ppim.book.....S}. The dust sticking coefficient $\xi_a$ varies with temperature for different elements/molecules \citep[e.g.,][]{2024A&A...692A.249B}. For the estimate at order of magnitude, we conservatively adopt $\xi_a=0.1$ in the calculation of the growth of dust grains, though $\xi_a\approx 1$ is suggested in \citet{1978ppim.book.....S}. Assuming pressure balance between the hot gas and the cold gas clumps formed in the rotating flow, we can estimate the growth timescale of the dust grains in the cold gas with solar abundance as
\begin{equation}
    t_{\rm dust}\sim {\frac {a_0}{{da}/{dt}}}\sim 3.74\times 10^{-10} a_0\alpha m\dot{m}_{\rm h}^{-1}r^{5/2}\tilde{H}_{\rm h} T_{\rm cl}^{1/2}~{\rm s}, \label{t_dust}
\end{equation}
where $a_0$ is the typical radius of dust grains.  Compare the growth timescale of the dust grains with the free-fall timescale (\ref{t_ff}), we have 
\begin{displaymath}
    {\frac {t_{\rm dust}}{t_{\rm ff}}}\sim 5.68\times 10^{-5}\alpha\tilde{H}_{\rm h} \left({\frac {a_0}{\rm 2\mu m}}\right)\left({\frac {\dot{m}_{\rm h}}{0.01}}\right)^{-1} ~~~~~~~~~~
\end{displaymath}
\begin{equation}
    \times\left({\frac {\lambda_{\rm d}}{0.01}}\right)^{1/2} \left({\frac {m}{10^8}}\right)^{-1/2} \left({\frac {T_{\rm cl}}{10^3~{\rm K}}}\right)^{1/2}, \label{t_dust_ff}   
\end{equation}
where $r=r_{\rm torus}$ is adopted. It is obvious that dust formation timescale is always much shorter than the free-fall timescale for any plausible values of the parameters for AGNs, which means the dust grains grow through accretion very rapidly in the cooling gas clumps.  }

\subsection{Radiation pressure supported torus}\label{sec:rad_torus}

The relative thickness of the dusty torus should be around the unity estimated from the ratio of type 1 to type 2 AGNs \citep[][]{1993ARA&A..31..473A}. The cold gas clumps in the dusty torus are irradiated by the emission from the accretion   disk in the AGN, and they re-radiate mainly in infrared wavebands. 
Although the detailed structure of the dusty torus is still quite unclear, some basic properties of the dusty torus can be worked out as follows.

The solid angle subtended by the whole dusty torus is 
\begin{equation}
    \Delta \Omega_{\rm torus}={\frac {4\pi H_{\rm torus}}{(R_{\rm torus}^2+H_{\rm torus}^2)^{1/2}}}, \label{Omega_torus}
\end{equation}
where $H_{\rm torus}$ is the half-thickness of the torus. If we assume that all the infrared emission of an AGN is from the dusty torus irradiated by the   disk emission, the infrared luminosity of the torus is
\begin{equation}
    L_{\rm ir}\sim {\frac {\Delta \Omega_{\rm torus}}{4\pi}}L_{\rm   disk}={\frac {H_{\rm torus}}{(R_{\rm torus}^2+H_{\rm torus}^2)^{1/2}}} L_{\rm   disk} \label{l_infrared}
\end{equation}
\citep*[][]{2005ApJ...619...86C}. 
The infrared flux of the dusty torus is then estimated by
\begin{displaymath}
   f_{\rm ir}\sim {\frac {L_{\rm ir}}{4\pi R_{\rm torus}\Delta R_{\rm torus}}} ~~~~~~~~~~~~~~~~~~~~~~~~~~~~~~~~~~~~~~~~~~~~~~~~
\end{displaymath}
\begin{equation}
={\frac {H_{\rm torus}}{4\pi R_{\rm torus}\Delta R_{\rm torus}(R_{\rm torus}^2+H_{\rm torus}^2)^{1/2}}} L_{\rm   disk}, \label{f_infrared}
\end{equation}
where the infrared flux is implicitly assumed to be equally emitted from the torus with a radial extension of $\Delta R_{\rm torus}$. The radial temperatures and emitted flux would decrease with distance, which are too complicated to be included in this work. However, we believe that general features of the dusty torus can be fairly well described by Equation (\ref{f_infrared}).

It was suggested that the dusty torus consists of clumpy clouds. In this work, we assume the vertical component of the BH gravity for the clumps in the inner region of the torus is balanced by the force due to the infrared radiation from the torus together with the UV/optical radiation of the   disk, we have
\begin{displaymath}
    {\frac {L_{\rm   disk}z}{4\pi (R_{\rm torus}^2+z^2)^{3/2}c}}
+{\frac {L_{\rm   disk}(1-e^{-\tau_{\rm cl}^{\rm ir}})z}{4\pi (R_{\rm torus}^2+z^2)^{1/2} R_{\rm torus}\Delta R_{\rm torus}c}}~~~~~~~~~~~~~~~~~~~~~~~~~~~~~~~~~~~~~~~~~
\end{displaymath}
\begin{equation}
~~~~~~~~~~~~~~~~~~~~={\frac {GM_{\rm bh}\Sigma_{\rm cl}z}{(R_{\rm torus}^2+z^2)^{3/2}}},\label{force_z}
\end{equation}
{since the clumps are always optically thick for UV photons,} where $\Sigma_{\rm cl}$ is the surface density of individual clumps, $\tau_{\rm cl}^{\rm ir}=\Sigma_{\rm cl}\kappa_{\rm ir}$. {The opacity $\kappa_{\rm ir}=6~{\rm cm^2~g}^{-1}$ for the infrared photons with temperature between $\sim 100-1000~{\rm K}$ \citep[][]{1985Icar...64..471P,2003A&A...410..611S}.  }
Letting $z=H_{\rm torus}$, we obtain
\begin{displaymath}
    {\frac {N_{\rm cl}}{10^{24}~{\rm cm}^{-2}}}=1.17~~~~~~~~~~~~~~~~~~~~~~~~~~~~~~~~~~~~~~~~~~~~~~~~~~~~~~~~~~~~~~~~~~~~~~~~~
\end{displaymath}
\begin{equation}
 \times\left[1+(1-e^{-\tau_{\rm cl}^{\rm ir}}) (1+\tilde{H}_{\rm torus}^2)(\Delta R_{\rm torus}/R_{\rm torus})^{-1}\right]\lambda_{\rm d},
    \label{lambda_d}
\end{equation}
from Equation (\ref{force_z}).

As the radial structure of the torus is still unclear, it is reasonable to approximate $\Delta R_{\rm torus}\sim R_{\rm torus}$, which is equivalent to the assumption that the most radiation from the   disk is blocked by the dusty clumps in the region with radial extention of $\Delta R_{\rm torus}\sim R_{\rm torus}$ in the torus. 
For the torus with $\Delta R_{\rm torus}\gg R_{\rm torus}$, the above estimate may be invalid, and one has to calculate the detailed radiation transfer in the torus, which is beyond the scope of this work. 

The relations of the Eddington ratio $\lambda_{\rm d}$ of the disk with the column density $N_{\rm cl}$ of the cold gas clumps are depicted with different values of $\tilde{H}_{\rm torus}$ in Figure \ref{fig:ncl_lambda}. The results are insensitive with the value of $\tilde{H}_{\rm torus}$. We note that the vertical structure of the torus is maintained solely by the radiation force, and no random motion of the clumps is required, which indicates our model does not suffer from too frequent collisions.

\begin{figure}
	\centering
	\includegraphics[width=0.9 \columnwidth]{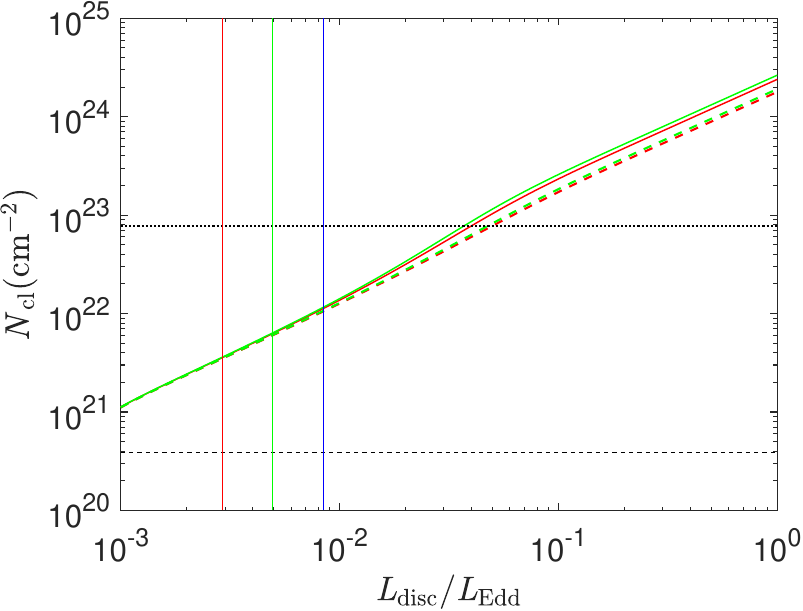}
	\caption{The relation of the column density $N_{\rm cl}$ of the cold gas clumps with the Eddington ratio $\lambda_{\rm d}$ of the  disk (see Equation \ref{lambda_d}). The red and green lines are the results for $\tilde{H}_{\rm torus}=0.2$ and $0.5$ respectively.  The solid lines indicate the results calculated with $\Delta R_{\rm torus}=R_{\rm torus}$, while the dashed lines are for $\Delta R_{\rm torus}=2R_{\rm torus}$. The black dashed and dotted lines indicate $\tau_{\rm cl}=1$ for the UV and infrared photons respectively.  The vertical lines indicate the critical Eddington ratios of the disk luminosity for thermal instability with different values of BH masses: $m=10^7$ (red), $10^8$ (green), and $10^9$ (blue). }
	\label{fig:ncl_lambda}
\end{figure}

\begin{figure}
	\centering
	\includegraphics[width=0.9 \columnwidth]{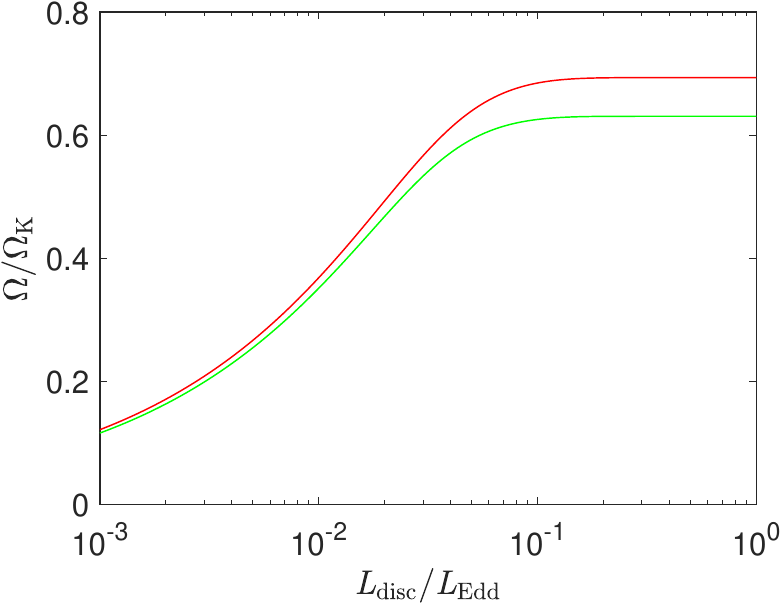}
	\caption{The circular angular velocities $\Omega$ vary with the Eddington ratio $\lambda_{\rm d}$. The red and green lines are the results for $\tilde{H}_{\rm torus}=0.2$ and $0.5$ respectively.   }
	\label{fig:lambda_omega_b}
\end{figure}

It is found that the cold gas clumps with rather low column density ($N_{\rm cl}\lesssim 10^{22}~{\rm cm}^{-2}$, still optically thick for UV photons) can be elevated vertically by the radiation force even if the disk luminosity $\lambda_{\rm d}$ is as low as $\sim 10^{-3}$, whereas the thermal instability is suppressed when $\lambda_{\rm d}\lesssim 3\times 10^{-3}-0.01$, which implies that no cold gas clumps cannot be formed in the case of low-$\lambda_{\rm d}$ (in the left side of the vertical lines in Figure \ref{fig:ncl_lambda}).  
Our model is naturally consistent with the lack of observational evidence of the dusty torus in some low-luminosity AGNs \citep[e.g.,][]{1999A&A...349...77C,2006ApJ...648L.101E,2008ARA&A..46..475H}. We note that, if the clumps have column density lower than the lines in Figure \ref{fig:ncl_lambda}, they will be driven away by the radiation force, while the high-$N_{\rm cl}$ clumps may sink towards the mid-plane of the torus. For typical AGNs with $\lambda_{\rm d}\sim 0.01-0.1$, the column density of the gas clumps in the radiation pressure supported torus should be in the range of $\sim 10^{22-23} {\rm cm}^{-2}$, which is consistent with the estimates of \citet{2007ApJ...661...52K} for a smooth torus. {The cold gas with column density $\sim 10^{22-23}~{\rm cm}^{-2}$ has been indeed detected in the circumnuclear region of Seyfert galaxies in the millimeter waveband \citep[][]{2018ApJ...867...48I,2021A&A...652A..98G}. A positive correlation has been found between the gas column densities responsible for the absorption of X-rays and the molecular gas column densities derived from CO toward the AGN \citep[][]{2021A&A...652A..98G}. Recent X-ray observations have also detected the clouds with similar column density $\sim 10^{22-23} {\rm cm}^{-2}$ \citep[][]{2023MNRAS.525.1941K,2023A&A...678A.154T,2025ApJ...981...91T}.}

For the clumps in circular orbits, their circular velocity is always sub-Keplerian due to the radiation force exerted on the clumps, which is 
\begin{equation}
    \tilde{\Omega}={\frac {\Omega}{\Omega_{\rm K}}}
    ={\frac {1}{(1+\tilde{H}_{\rm torus}^2)^{3/4}}}
    \left[1-{\frac {\lambda_{\rm d}}{8.54 \times 10^{-25}N_{\rm cl}}}\right]^{1/2},\label{omega_0}
\end{equation}
 {derived from the radial component of the momentum equation.} The column density $N_{\rm cl}$ of the clumps as a function of $\lambda_{\rm d}$ is derived with Equation (\ref{lambda_d}) when $\tilde{H}_{\rm torus}$ is specified. Use Equation (\ref{omega_0}), we plot the results of the angular velocity of the clumps varying with the Eddington ratio $\lambda_{\rm d}$ in  Figure \ref{fig:lambda_omega_b}.

\begin{figure}
	\centering
	\includegraphics[width=0.9 \columnwidth]{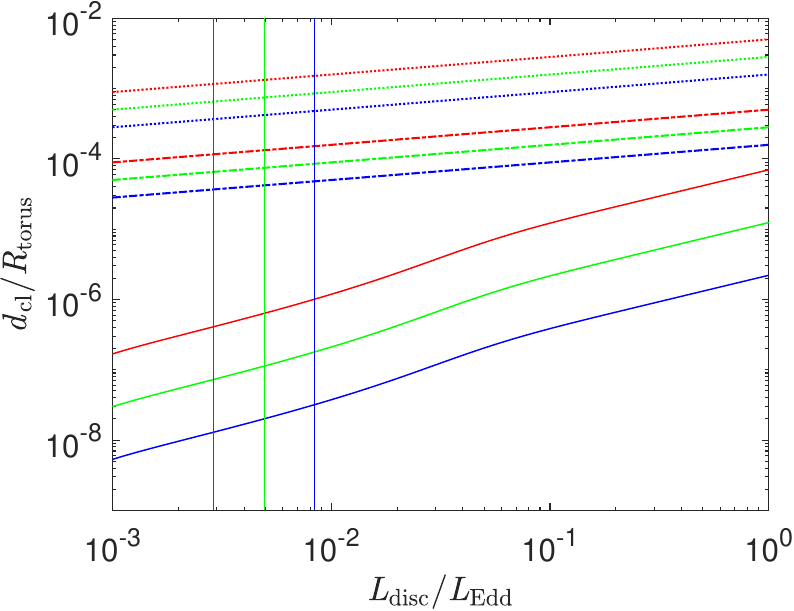}
	\caption{The sizes of the clumps vary with the Eddingtion ratio $\lambda_{\rm d}$ of the disk emission. The parameters, $\alpha=0.1$, $\tilde{H}_{\rm h}=0.5$, and $\dot{m}_{\rm h}=\lambda_{\rm d}$ are adopted in the calculations with Equation (\ref{rcl}) (solid lines). The colored lines indicate the results with different values of black hole mass, $m=10^7$ (red), $10^8$ (green), and $10^8$ (blue), respectively. The vertical lines indicate the critical Eddington ratios of the disk luminosity for thermal instability. The dotted lines indicate the minimal size of the sinking clumps required to avoid dissolved due to the Kelvin-Helmholtz instability calculated with Equation (\ref{rclr_kh_inst2}) ($z_{\rm i}=H_{\rm h}$), whereas the dash-dotted lines are the results calculated for $z_{\rm i}=0.1 H_{\rm h}$.}  
	\label{fig:lambda_rcl_b}
\end{figure}

\begin{figure}
	\centering
	\includegraphics[width=0.9 \columnwidth]{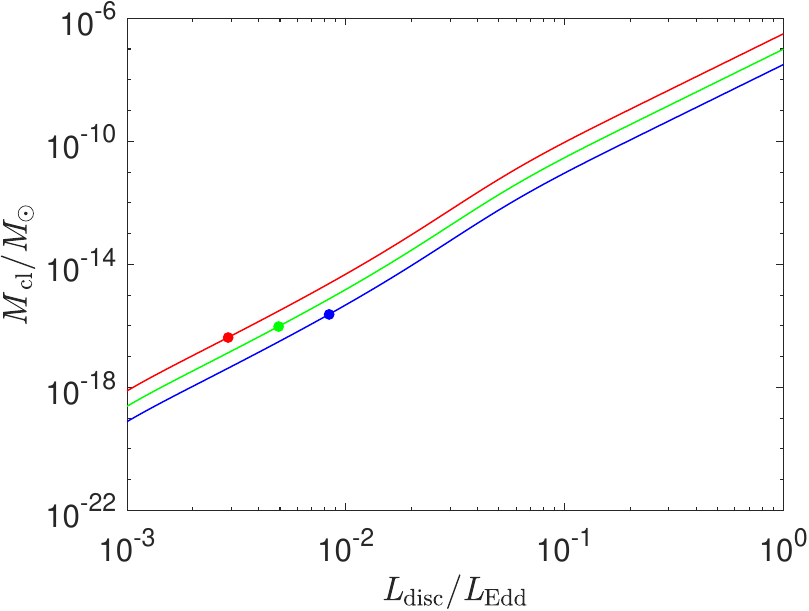}
	\caption{The same as Figure \ref{fig:lambda_rcl_b}, but for the mass of the clump varying with the Eddingtion ratio of the disk emission calculated with Equation (\ref{mcl}).  }
	\label{fig:lambda_mcl_b}
\end{figure}

\subsection{Structure of the clumps}\label{sec:stru_clumps}

For the clumps with low column density, the self-gravity is always negligible. We assume that the cold clumps are in pressure equilibrium with the ambient hot gas. The dust in the clumps will be sublimated if the temperature is higher than $\sim 1500$~K. The clumps are irradiated by the central   disk emission and re-radiate infrared photons. Thus, the temperature of the clumps is mainly determined by the collective radiation of the clumps in the torus. It should be lower than the sublimation temperature, and we assume $T_{\rm cl}=1000$~K in all our calculations. The properties of the hot gas varying with radius can be calculated with Equations (\ref{v_r_h})-(\ref{rho_h}). Assuming the cold gas  clumps to be in pressure equilibrium with the ambient hot gas, $p_{\rm cl}=p_{\rm h}$, we obtain the density $\rho_{\rm cl}$ of the cold clumps. We finally derive the size and the mass of the clumps as functions of $\lambda_{\rm d}$, 
\begin{displaymath}
    {\frac {d_{\rm cl}}{R_{\rm torus}}}=2.98\times 10^{-4}\alpha {\tilde H}_{\rm h}
    \left({\frac m {10^8}}\right)^{-3/4}
    \left({\frac {\dot{m}_{\rm h}}{0.01}}\right)^{-1}~~~~~~~~~~~~~~~~~~~~~~~~~~~~~~~~~
\end{displaymath}
\begin{equation}
    \times\left({\frac {\lambda_{\rm d}}{0.01}}\right)^{3/4}\left({\frac {T_{\rm cl}}{10^3}}\right) \left({\frac {N_{\rm cl}}{10^{24}}}\right), \label{rcl}
\end{equation}
and
\begin{displaymath}
    {\frac {M_{\rm cl}}{M_\odot}}=2.15\times 10^{-7}  \alpha^2{\tilde H}_{\rm h}^2
    \left({\frac m {10^8}} \right)^{-1/2} 
    \left({\frac {\dot{m}_{\rm h}}{0.01}}\right)^{-2}~~~~~~~~~~~~~~~~~~~~~~~~~~~
\end{displaymath}
\begin{equation}
    \times\left({\frac {\lambda_{\rm d}}{0.01}}\right)^{5/2}
    \left({\frac {T_{\rm cl}}{10^3}}\right)^2
    \left({\frac {N_{\rm cl}}{10^{24}}}\right)^3, \label{mcl}
\end{equation}
respectively, in which the column density $N_{\rm cl}$ as a function of the Eddinton ratio of the   disk $\lambda_{\rm d}$ is given in Equation (\ref{lambda_d}). As discussed in Section \ref{sec:condense}, the hot gas may condense to the cold clumps only if the Eddington ratio of the disk $\lambda_{\rm d}\gtrsim 3\times 10^{-3}-0.01$ (see Figure \ref{fig:arate_h_crit}). 
The results of the clump size and mass varying with the Eddington radio $\lambda_{\rm d}$ are plotted in Figures \ref{fig:lambda_rcl_b} and \ref{fig:lambda_mcl_b}, respectively. We find that the clump sizes are always much larger than the critical wavelength $\lambda_{\perp,\rm cr}$ required for thermal instability (see Figure \ref{fig:wavelenth_crit}).

\subsection{Structure of the dusty torus}\label{sec:stru_torus}

Although we do not intend to explore the detailed spatial distribution of clumps in the torus in this work, the space density $n_{\rm clump}$ of the cold gas clumps can be constrained by 
\begin{equation}
\tau_{\rm torus}\sim n_{\rm clump}\pi R_{\rm cl}^2 \Delta R_{\rm torus}>1, \label{tau_torus} 
\end{equation}
in which we have implicitly assumed the individual clump to be optically thick for the disk radiation of the AGN, and $R_{\rm cl}\approx d_{\rm cl}/2$ is the radius of the clumps. The mean free path for collisions between clumps is 
\begin{equation}
    l_{\rm colli}={\frac 1 {n_{\rm clump}\pi R_{\rm cl}^2}}\sim {\frac {\Delta R_{\rm torus}}{\tau_{\rm torus}}}, \label{l_colli}
\end{equation}
where Equation (\ref{tau_torus}) is used, which is comparable with ${H}_{\rm torus}$ if $\Delta R_{\rm torus}\sim R_{\rm torus}$ and $\tau_{\rm torus}$ is not much greater than the unity.

{When a heavy cold clump is formed in the hot gas flow, the radiation force is unable to counter the vertical gravity of the BH, it will sink down to the mid-plane of the disk. The clump is accelerated by the vertical component of the BH  gravity, and we have
\begin{equation}
    {\frac {dv_z}{dt}}\sim {\frac {GM_{\rm bh}z}{R^3}}, \label{dvzdt}
\end{equation}
where a simplified vertical gravity slightly different from that given in Section \ref{sec:rad_torus} is used. It should be sufficiently accurate for the estimate of the sink timescale. Integrating Equation (\ref{dvzdt}), we obtain 
\begin{equation}
    v_z(z)={\frac {dz}{dt}}\sim v_{\rm K}\left [  
 \left({\frac {z_{\rm i}}{R}}\right)^2-\left({\frac {z}{R}}\right)^2
        \right]^{1/2}, \label{dzdt}
\end{equation}
where the clump is assumed to be formed at $z=z_{\rm i}$. The maximal velocity $v_{z,\rm max}=v_{\rm K}z_{\rm i}/R$ when the clump approaches the mid-plane. The sink time, i.e., the time of the clump moves from $z=z_{\rm i}$ to $z=0$, is derived by integrating Equation (\ref{dzdt}),
\begin{equation}
    t_{\rm sink}\sim {\frac {\pi}{2\Omega_{\rm K}}}, \label{t_sink}
\end{equation}
which is independent of $z_{\rm i}$, and is at the same order of the dynamical timescale. It implies that sufficiently heavy clumps may on average experience one collision with one floating clump within the sink timescale when they are falling down to the mid-plane, as the mean free path $l_{\rm colli}\sim {H}_{\rm torus}$ (see Equation \ref{l_colli}). However, the cold gas clumps confined by the pressure of ambient hot gas are subject to Kelvin-Helmholtz (KH) instability if the clumps are moving through the hot gas at a large relative velocity, which may dissolve the cold clumps \citep[][]{1993ApJ...407..588M}.} It was suggested that the self-gravity of the clumps may suppress the instability, while the column density of the cold clumps in the torus is rather low, of which the self-gravity is negligible. {The analysis of the KH instability is given in Appendix \ref{analysis_KH_inst}.} 


The constraints on the clump size by the KH instability calculated with Equation (\ref{rclr_kh_inst2}) are plotted in Figure \ref{fig:lambda_rcl_b}, in which the maximal relative velocity $U=v_{\rm K}z_{\rm i}/R$ is adopted. 
{It is found that the clumps with size $d_{\rm cl}\lesssim 10^{-3}R_{\rm torus}$ in the geometrically thick torus must be dissolved during their free falling onto the mid-plane, while the clumps with proper column density staying statically above/below the mid-plane in the torus are immune to the KH instability due to zero (or very small) relative velocity $U$. We note that, even for the clumps in vertical equilibrium, their sizes are much smaller than the critical size given by Equation (\ref{rclr_kh_inst2}) (see Figure \ref{fig:lambda_rcl_b}). They may be dissolved if there is any perturbation makes them move through ambient hot gas at a large relative velocity, which will be discussed in Section \ref{sec:stability_torus}. }

\subsection{Stability of the dusty torus}\label{sec:stability_torus}

The hot gas may condense to cold clumps when the accretion rate of the hot flow is higher than the critical value (see the discussion in Section \ref{sec:condense}). {Only the clumps with suitable column density can be in the vertical equilibrium, and the remainder are either driven away or sinking down. The heavier clumps will soon be dissolved due to the KH instability. Even for those in vertical static state, a small perturbation may lead to disruption of the clumps, which need to be refilled by the clumps condensed from the hot gas, otherwise the dusty torus will soon disappear. Let us now turn to the issue of the dynamical stability and robustness of the
clumps that may form out of the hot gas to build a clumpy dusty torus.}      

{The production rate of the clumps in unit volume is estimated as
\begin{equation}
    \dot{n}_{\rm clump}^+\sim {\frac {f_{\rm cond}\rho_{\rm h}}{M_{\rm cl}t_{\rm cool}}},\label{dn_clumpdt+}
\end{equation}
where the parameter $f_{\rm cond}$ describes the efficiency of the hot gas condensed to the clumps in vertical static equilibrium. We note that the value of $f_{\rm cond}$ is very uncertain due to ignorance of the detailed physics of the gas condensation, though $f_{\rm cond}\lesssim1$ is generally required. }

{The dissipation rate of the clumps is
\begin{equation}
   \dot{n}_{\rm clump}^-\sim {\frac {n_{\rm clump}}{t_{\rm dissi}}}={\frac {n_{\rm clump}}{f_{\rm dissi}t_{\rm sink}}},     \label{dn_clumpdt-}
\end{equation}
where the dissipation timescale $t_{\rm dissi}$ is assumed to scale with the sink time as $t_{\rm dissi}=f_{\rm dissi}t_{\rm sink}$. Thus the time evolution of the torus is described by
\begin{equation}
    {\frac {dn_{\rm clump}}{dt}}=\dot{n}_{\rm clump}^+-\dot{n}_{\rm clump}^-. \label{evol_n_clump}
\end{equation}
In the case of $\dot{n}_{\rm clump}^+>\dot{n}_{\rm clump}^-$, $dn_{\rm clump}/{dt}>0$, the space number density of the clumps increases with time till $\dot{n}_{\rm clump}^+=\dot{n}_{\rm clump}^-$, and vice verse. A steady clumpy torus is therefore maintained through this mechanism.} The structure of a steady torus can be obtained by letting $\dot{n}_{\rm clump}^+=\dot{n}_{\rm clump}^-$, i.e., 
\begin{equation}
    n_{\rm clump}R_{\rm torus}^3=f_{\rm dissi}f_{\rm cond}R_{\rm torus}^3{\frac {\rho_{\rm h} t_{\rm sink}}{M_{\rm cl}t_{\rm cool}} }.\label{n_clump_2}
\end{equation}
For the clumps in vertical force equilibrium, their column density and size vary with the   disk luminosity if the values of all other model parameters are specified (see Equations \ref{lambda_d} and \ref{rcl}). We plot the space number density of the clumps as a function of the Eddington ratio of the disk emission in Figure \ref{fig:lambda_nclump}.

\begin{figure}
	\centering
	\includegraphics[width=0.9 \columnwidth]{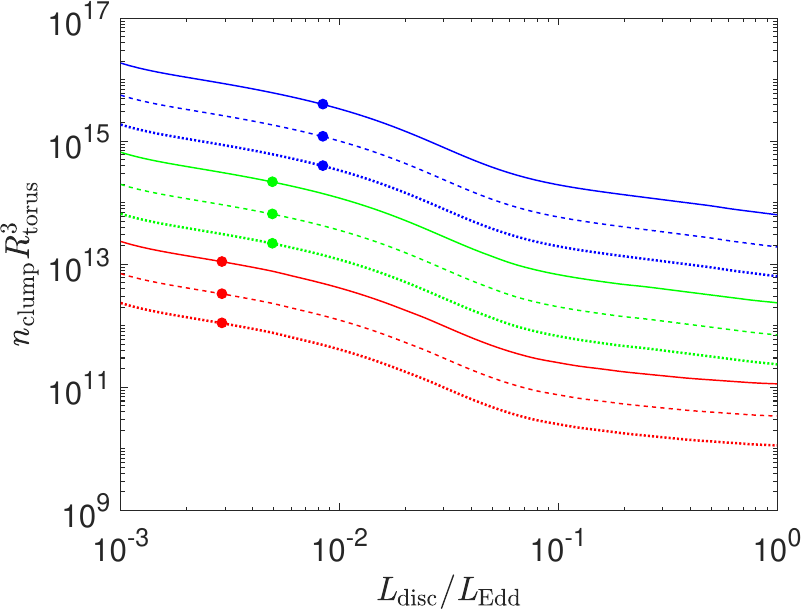}
	\caption{The space density of the clumps in a steady torus varies with the Eddington ratio of the disk emission calculated with Equation (\ref{n_clump_2}). The parameters, $\alpha=0.1$, $\tilde{H}_{\rm h}=0.5$, $f_{\rm dissi}=1$, and $\dot{m}_{\rm h}=\lambda_{\rm d}$ are adopted in the calculations. The colored lines indicate the results with different values of black hole mass, $m=10^7$ (red), $10^8$ (green), and $10^9$ (blue), respectively. The different types of the lines are for the results with $f_{\rm cond}=1$ (solid), $0.3$ (dashed), and $0.1$ (dotted). The dots indicate the critical Eddington ratios of the disk luminosity.}
	\label{fig:lambda_nclump}
\end{figure}

\begin{figure}
	\centering
	\includegraphics[width=0.9 \columnwidth]{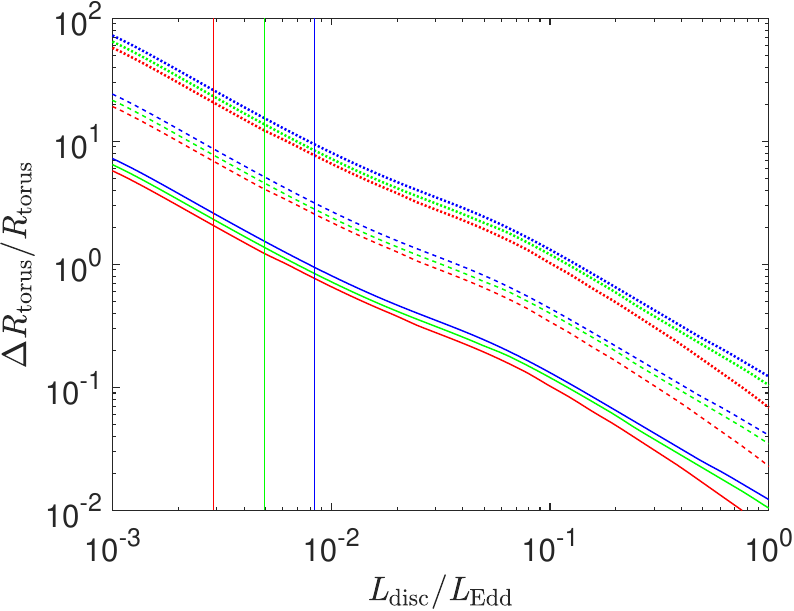}
	\caption{The radial extension of the torus varies with the Eddington ratio of the   disk emission calculated with Equation (\ref{dr_torus_r_torus}). The parameters, $\alpha=0.1$, $\tilde{H}_{\rm h}=0.5$, $f_{\rm dissi}=1$, and $\dot{m}_{\rm h}=\lambda_{\rm d}$ are adopted in the calculations. The different types of the lines are for the results with $f_{\rm cond}=1$ (solid), $0.3$ (dashed), and $0.1$ (dotted). The colored lines indicate the results with different values of black hole mass with different values of BH masses: $m=10^7$ (red), $10^8$ (green), and $10^9$ (blue). The vertical lines indicate the critical Eddington ratios of the disk luminosity for thermal instability.  }
	\label{fig:lambda_dr_torus}
\end{figure}

With derived space number $n_{\rm clump}$, we can estimate the radial extension of the torus using Equation (\ref{tau_torus}),
\begin{equation}
    {\frac {\Delta R_{\rm torus}}{R_{\rm torus}}}\sim {\frac {\tau_{\rm torus}}{\pi}}\left( 
    {\frac {R_{\rm cl}}{R_{\rm torus}}}\right)^{-2}(n_{\rm clump}R_{\rm torus}^3)^{-1},\label{dr_torus_r_torus}
\end{equation}
if the values of the parameters are given. In the estimate of $\Delta R_{\rm torus}$, we adopt $\tau_{\rm torus}=3$, which roughly corresponds to $\sim 95\%$ of the   disk emission within the solid angle subtended by the torus is absorbed by the clumps in the torus. We note that the results are independent of the BH mass. The radial extension of the torus $\Delta R_{\rm torus}$ varying with the Eddington ratio of the   disk emission is plotted in Figure \ref{fig:lambda_dr_torus}.

\section{Discussion}\label{sec:discussion}

The hot gas will condense to the cold clumps due to thermal instabilities in the rotating hot gas flow only if $\dot{m}_{\rm h}\gtrsim 0.001-0.01$ (see Figure \ref{fig:arate_h_crit}). These cold gas clumps in the inner region are irradiated by the emission of the disk surrounding the black hole (BH). The clumps with suitable column density are therefore elevated away from the mid-plane by the infrared radiation together with the vertical component of the disk radiation force. {During the cooling of the clumps, the dust can be formed rapidly due to high density of the clumps,} which may appear collectively as dusty torus blocking part of the photons emitted from the disk. Our model calculations show that, the thermal instability is suppressed when the accretion rate of the hot gas is lower than this critical rate, so no cold gas can be condensed from the hot gas in the flow. These are consistent with the observations showing the absence of the obscuring matter in low-luminosity AGNs \citep[e.g.,][]{1999A&A...349...77C,2006ApJ...648L.101E,2008ARA&A..46..475H,2011ApJ...733...60T,2015A&A...578A..74G,2022MNRAS.510.5102O}. In this case, the hot gas feeds the BH most probably through an ADAF \citep[][]{1994ApJ...428L..13N}, which is believed to take place in the low-luminosity AGNs.

We expect the column density distribution of the cold clumps condensed from the hot gas may spread in a certain range, though the details are almost unknown, as the physics of the non-linear growth of the thermal instabilities is still quite uncertain \citep[see][for review and the references therein]{2023FrASS..1098135W}. 

{Our calculations show that only the clumps with suitable column density $N_{\rm cl}$ can be elevated above/below the mid-plane in vertically quasi-static balance between the gravity and radiation force, which mean that the clumps with higher $N_{\rm cl}$ would sink to the mid-plane, because they are too heavy to be supported vertically by the radiation force. These clumps suffer from the KH instability due to their large relative velocity with the ambient hot gas, and they are dissolved when moving through the ambient hot gas. Only those clumps in vertically quasi-static equilibrium can be immune to the KH instability. Thus, the KH instability plays a role as a filter to pick out   cold clumps condensed from the hot gas to appear collectively as a dusty torus.} 

The clumps condensed from the hot gas in the region very close to the mid-plane with zero or very small vertical velocities may survive through the KH instability, which implies that the cold clumps may be accumulated in this region. Some of these may leave inwards from the torus, and finally fall onto the accretion  disk to feed the BH, of which the detailed physics processes are beyond the scope of this work. We speculate that some extremely light (low-$N_{\rm cl}$) clumps may be driven away from the torus by the radiation force. Some of them may survive through the KH instability, which may be the candidates appear as the dust observed in the polar region of some AGNs \citep[][]{2012ApJ...755..149H,2013ApJ...771...87H,2014A&A...563A..82T,2014A&A...565A..71L,2022A&A...667A.140G}.

For clumps with suitable column density, the vertical component of the BH gravity is in quasi-static equilibrium with the infrared radiation force together with the vertical component of the   disk radiation force, which implies its vertical velocity is very small close to nil. In our model, the trajectories of most individual clumps are vertically layered. Thus, the frequent collisions between chaotic moving clumps in the torus in the previous works can be avoided. We note that, even for those in vertical static state, a small density perturbation to an individual clump may alter its dynamic equilibrium. Thus, we expect that, overall, the the clumps of suitable column densities are only in quasi-equilibrium state. A balance between the finite time of this quasi-equilibrium and clump production time scale by cooling off of the hot gas determines the overall number density of the clumps and the radial extension of the torus of optical depth equal to unity. In the case of clump production rate is larger than the clump destruction rate, the space number density of the clumps increases with time until a balance between them is attained, and vice verse.
We showed that for any reasonable time scale of the quasi-static equilibrium our conclusion with respect to the existence of a dynamic torus of the kind proposed here is robust (see Section \ref{sec:stability_torus}).

The geometrically thick torus is maintained by the radiation force, which is governed predominantly by the emission of the accretion   disk. We find that no geometrically thick torus is present if the   disk luminosity is low ($\lambda_{\rm d}\lesssim 0.001$). In this case, one may expect the clumps with extremely low column density to be driven by the radiation force, however, such low column density makes the clumps even optically thin, which substantially reduces the radiation force exerted on the clumps. Thus, the radiation force is unable to overcome the vertical component of the gravity of the clumps, and most of them may sink to the mid-plane. The situation is similar in the region beyond the torus, where the condensed cold clumps should sink downwards rapidly due to lack of radiation force to counter the gravity in the vertical direction. As discussed in Sections \ref{sec:stru_torus} and \ref{sec:stability_torus}, only a small fraction of the clumps formed in the region close to the mid-plane may survive through the KH instability.  

The outer radius of the dusty torus can be estimated where most of the   disk emission is absorbed by the clumps in the solid angle subtended by the torus. We find that the relative radial extension of the torus varies over several orders of magnitude. We predict that the radial extension of the torus decreases with increasing Eddington rate. 
It should be cautious that the results in Figure \ref{fig:lambda_dr_torus} are only qualitative, because the column density of the clumps derived with Equation (\ref{lambda_d}) is only valid for $\Delta R_{\rm torus}\sim R_{\rm torus}$, otherwise, one has to solve a two-dimensional radiation transfer equation, which is beyond the scope of this work.    

There are plenty of observational evidence of hot gas in the central region of the galaxies, which is supposed to be the reservoir to feed the central massive BH. If the accretion rate of the hot gas is higher than the critical value, cold clumps form through condensing from the hot gas in the flow, part of which is puffed up vertically by the radiation force as the dusty torus, while the remainder may fall onto the BH through a cold   disk, namely,  the ``cold accretion" in normal bright AGNs/quasars. In the low-luminosity AGNs or normal galaxies including our own galaxy, the Bondi accretion rate is too low to make the hot gas cool down, and therefore an ADAF may surround the BH without a dusty torus. Such accretion of hot gas is usually named as ``hot accretion" in order to distinguish from ``cold accretion" in bright AGNs/quasars \citep[][]{2007MNRAS.376.1849H}.

\begin{acknowledgments}
We are grateful to the referee for his/her insightful comments and suggestions. We thank Aigen Li, Tinggui Wang, Luis Ho, Bożena Czerny, Alex Markowitz and Junfeng Wang for helpful discussion. This work is supported by the NSFC ((12533005, 12233007, and 12347103), the science research grants from the China Manned Space Project with No. CMS-CSST-2025-A07, and the fundamental research fund for Chinese central universities (Zhejiang University).
\end{acknowledgments}



\bibliographystyle{apsrev4-2}
\bibliography{dust_torus}
\appendix

\section{Linear analysis on thermal instability}\label{analysis_thermal_inst}


{It is well known that the hot gas always suffers from thermal instability without considering thermal conduction in the gas, while the perturbations with wavenumbers greater than a critical wavenumber $k_{\rm cr}$ under certain circumstance can be stabilized through thermal conduction \citep[][]{1965ApJ...142..531F}. Here we apply Field's linear analysis on the thermal instability for the magnetized hot gas flow. The critical wavenumber perpendicular to the magnetic field line is
\begin{displaymath}
    k_{\perp,\rm cr}=k_{\rm cond,\perp}^{1/2}\left[k_\rho\left( 1+ {\frac {v_{\rm A,h}^2}{c_{\rm s,h}^2}}\right)^{-1}-k_T\right]^{1/2}   
\end{displaymath}
\begin{equation}   
 ~~~~~~~~~  =k_{\rm cond,\perp}^{1/2}\left[k_\rho\left( 1+ {\frac {2}{\beta_{\rm h}}}\right)^{-1}-k_T\right]^{1/2}, \label{k_c}
\end{equation}
where $\beta_{\rm h}=8\pi p_{\rm h}/B_{\rm h}^2$, the Alfven speed $v_{\rm A,h}=B_{\rm h}/(4\pi\rho_{\rm h})^{1/2}$, the wavenumbers, $k_\rho$, $k_T$, and $k_{\rm cond,\perp}$, are given by
\begin{equation}
k_\rho={\frac {\mu m_{\rm p}}{k_{\rm B}}} {\frac {(\gamma-1)f_{\rm rad}^{-}} {\gamma^{1/2}c_{\rm s,h}\rho_{\rm h}T_{\rm h}}},\label{k_rho}
\end{equation}
\begin{equation}
k_T={\frac {\mu m_{\rm p}}{k_{\rm B}}} {\frac {(\gamma-1)\partial f_{\rm rad}^{-}/\partial T_{\rm h}} {\gamma^{1/2}c_{\rm s,h}\rho_{\rm h}}},\label{k_T}
\end{equation}
\begin{equation}
k_{\rm cond,\perp}={\frac {k_{\rm B}}{\mu m_{\rm p}}} {\frac {\gamma^{1/2}c_{\rm s,h}\rho_{\rm h}}{(\gamma-1)\kappa_{\perp}}},\label{k_cond_perp}
\end{equation}
and $\kappa_{\perp}$ is the conductivity normal to the magnetic field \citep[][]{1965ApJ...142..531F}. Here we assume the heating of the gas remains unchanged during a perturbation in the hot gas, which is the same as the previous analyses on the accretion flows \citep[][]{1983ApJ...267...18K,2013ApJ...767..156M}. In the presence of a magnetic field, the conduction in the gas is strongly reduced in the direction perpendicular to the field line, which reads
\begin{equation}
   {\frac {\kappa_{\perp}} {\kappa_{//}}}=4.8\times 10^{-10}n_{\rm h,e}^2B_{\rm h}^{-2}T_{\rm h}^{-3}=3.67\times 10^{28}\beta_{\rm h}\rho_{\rm h}T_{\rm h}^{-4},\label{kappa_ratio} 
\end{equation}
where the conductivity along the field line $\kappa_{//}=\kappa_0/3$, for a fully ionized gas with ion-electron collision timescale is much longer than the gyro-period \citep[][]{1962pfig.book.....S}, which is valid for our present discussion of a magnetized accretion   disk. The discussion for the general case can be found in \citet{1958JETP....6..358B}. The normal thermal conductivity without magnetic field $\kappa_0=10^{-6}T_{\rm h}^{5/2}~{\rm g~cm~K^{-1/2}~s^{-3}}$ \citep[][]{1962pfig.book.....S}.}

Substitute Equations (\ref{temp_h}), (\ref{rho_h}), (\ref{r_in/r_g}), and (\ref{kappa_ratio}) into Equation (\ref{k_c}), we obtain the critical wavelength of the perturbation at $R_{\rm torus}$ as $\lambda_{\perp,\rm cr}=2\pi/{k_{\perp,\rm cr}}$.

\section{Analysis on Kelvin-Helmholtz instability of clumps}\label{analysis_KH_inst}

The frequency of the KH mode is given by 
\begin{equation}
    \omega=\left[-{\frac {\rho_{\rm h}\rho_{\rm cl}k^2U^2}{(\rho_{\rm h}+\rho_{\rm cl})^2}}\right]^{1/2}, \label{omega}
\end{equation}
in the hydrodynamic analysis, where $U$ is the relative velocity between the clumps and the ambient hot gas \citep{1961hhs..book.....C}. It indicates that the KH instability is always present whenever there is a non-zero relative velocity. We note that the frequency becomes
\begin{equation}
    \omega=\left[  {\frac {k^2(B_{\rm h}^2+B_{\rm cl}^2)}{4\pi (\rho_{\rm h}+\rho_{\rm cl})}}
    -{\frac {\rho_{\rm h}\rho_{\rm cl}k^2U^2}{(\rho_{\rm h}+\rho_{\rm cl})^2}}\right]^{1/2}, \label{omega_b}
\end{equation}
in the presence of magnetic field \citep{1961hhs..book.....C}, where $B_{\rm h}$ and $B_{\rm cl}$ are the field strengths of the hot gas and the cold clumps respectively. For a clump moves downwards vertically, the  
KH instability criterion is 
\begin{displaymath}
   \omega^2= {\frac {k^2(B_{{\rm h},z}^2+B_{{\rm cl},z}^2)}{4\pi (\rho_{\rm h}+\rho_{\rm cl})}}
    -{\frac {\rho_{\rm h}\rho_{\rm cl}k^2U^2}{(\rho_{\rm h}+\rho_{\rm cl})^2}}  ~~~~~~~~~~~~~~~~~~~
\end{displaymath}
\begin{equation} 
    ={\frac {k^2(B_{{\rm h},z}^2+B_{{\rm cl},z}^2)}{4\pi (\rho_{\rm h}+\rho_{\rm cl})}}
    -{\frac {\rho_{\rm h}\rho_{\rm cl}k^2v_{\rm K}^2z_{\rm i}^2}{(\rho_{\rm h}+\rho_{\rm cl})^2R^2}}<0,
    \label{kh_inst}
\end{equation}
where $B_{{\rm h},z}$ and $B_{{\rm cl},z}$ are the vertical field strengths of the hot gas and the clumps respectively, and $U=v_{z,\rm max}=v_{\rm K}z_{\rm i}/R$ is adopted. The instability criterion (\ref{kh_inst}) can be re-written as 
\begin{equation}
   \omega^2=k^2c_{\rm s,cl}^2\left[ {\frac 2{\beta_{{\rm h},z}}}+{\frac 2{\beta_{{\rm cl},z}}}-\left({\frac {z_{\rm i}}{H_{\rm h}}}\right)^2\right]<0, \label{kh_inst2}
\end{equation}
or
\begin{equation}
    {\frac 2{\beta_{{\rm h},z}}}+{\frac 2{\beta_{{\rm cl},z}}}<\left({\frac {z_{\rm i}}{H_{\rm h}}}\right)^2, \label{kh_inst3}
\end{equation}
where $c_{\rm s,cl}=(p_{\rm cl}/\rho_{\rm cl})^{1/2}$, $\beta_{{\rm h},z}=8\pi p_{\rm h}/B_{{\rm h},z}^2$, $\beta_{{\rm cl},z}=8\pi p_{\rm h}/B_{{\rm cl},z}^2$, $H_{\rm h}/R=c_{\rm s,h}/v_{\rm K}$, and the approximation $\rho_{\rm cl}+\rho_{\rm h}\simeq \rho_{\rm cl}$ are used. It is well known that the vertical field strength of a hot accretion flow is always very low, i.e., $\beta_{{\rm h},z}\gg 1$, except its inner region where the gas is magnetically arrested under certain circumstance \citep{2003ApJ...592.1042I,2003PASJ...55L..69N}. As the cold clumps are condensed from the hot gas due to thermal instabilities, one may expect the field of the clumps could be stronger than that of the hot gas due to the freezing effect of magnetic field flux, though the detailed physics process is quite uncertain. Equation (\ref{kh_inst3}) shows that the KH instability is suppressed, if  
\begin{equation}
     \beta_{{\rm cl},z}<\beta_{{\rm cl},z}^{\rm cr}=2\left({\frac {z_{\rm i}}{H_{\rm h}}}\right)^{-2}. \label{kh_st}
\end{equation}
For $z_{\rm i}=H_{\rm h}$, $\beta_{{\rm cl},z}^{\rm cr}=2$, which means that the sinking clumps always suffer from the KH instability, unless the magnetic pressure is comparable with the gas pressure of the clumps or the clumps are formed in the region very close to the mid-plane.

Setting $k=2\pi/d_{\rm cl}$, the growth time of the KH instability is estimated as
\begin{equation}
  t_{\rm KH}\sim{\frac {2\pi}{\omega}}={\frac {d_{\rm cl}}{c_{\rm s,cl}}}\left({\frac {z_{\rm i}}{H_{\rm h}}}\right)^{-1},
    \label{t_kh}
\end{equation}
if the vertical magnetic field of the clumps is not very strong, i.e., $\beta_{{\rm cl},z}\gg 1$. Compare it with the sink time (\ref{t_sink}), we have
\begin{equation}
    {\frac {t_{\rm KH}}{t_{\rm sink}}}={\frac 2{\pi}}\left({\frac {c_{\rm s,cl}}{v_{\rm K}}} \right)^{-1}\left({\frac {z_{\rm i}}{H_{\rm h}}}\right)^{-1}{\frac {d_{\rm cl}}R}.
\end{equation}
The cold clumps condensed from the hot gas due to the thermal instabilities with column density higher than that given by Equation (\ref{lambda_d}) will sink downwards. The clumps with $t_{\rm KH}\lesssim t_{\rm sink}$, i.e., 
\begin{equation}
 {\frac {d_{\rm cl}}R}\lesssim {\frac {d_{\rm cl}^{\rm cr}}R}= {\frac {\pi}2}\left({\frac {c_{\rm s,cl}}{v_{\rm K}}} \right)\left({\frac {z_{\rm i}}{H_{\rm h}}}\right),
    \label{rclr_kh_inst}
\end{equation}
will be dissolved while they are sinking down to the mid-plane due to the KH instability. Substitute  Equation (\ref{r_in/r_g}) into Equation (\ref{rclr_kh_inst}), we have
\begin{displaymath}
    {\frac {d_{\rm cl}}R}\lesssim {\frac {d_{\rm cl}^{\rm cr}}R}=8.9\times 10^{-4} 
\left({\frac m {10^8}}\right)^{-1/4}~~~~~~~~~~~~~~~~~~~~~~~~~~~~~~~~~~~~~
\end{displaymath}
\begin{equation}
~~~~~~~~\times 
\left({\frac  {\lambda_{\rm d}}{0.01}} \right)^{1/4}
\left({\frac {T_{\rm cl}}{10^3~{\rm K}}}\right)^{1/2}
\left({\frac {z_{\rm i}}{H_{\rm h}}}\right),
   \label{rclr_kh_inst2} 
\end{equation}
when $R=R_{\rm torus}$.

For the cold clumps with suitable column density vertically supported by the radiation force, they are co-rotating with the hot gas   disk, whereas the relative radial velocity $U\simeq |v_{R,\rm h}|$ between clumps and hot gas may also be subject to the KH instability. The field of the clumps could be stronger than that of the hot gas. As we know that the magnetic field may stabilize the cold clumps in the flowing hot gas, we conservatively assume $B_{{\rm cl},r}=B_{{\rm h},r}$ in the instability analysis. We re-write Equation (\ref{omega_b}) as
\begin{equation}
    \omega=kc_{\rm s,cl}\left[{\frac {4}{\beta_{{\rm h},r}}}-\alpha^2\left({\frac {H_{\rm h}}{R}}\right)^2\right]^{1/2}, \label{omega_b2}
\end{equation}
where $c_{\rm s,cl}=(p_{\rm cl}/\rho_{\rm cl})^{1/2}$, $\beta_{{\rm h},r}=8\pi p_{\rm h}/B_{{\rm h},r}^2$, and the approximation $\rho_{\rm cl}+\rho_{\rm h}\simeq \rho_{\rm cl}$ is used. We find that the Kelvin-Helmholtz instability is suppressed if the condition, 
\begin{equation}
    \beta_{{\rm h},r}\le \beta_{{\rm h},r}^{\rm cr}=4\alpha^{-2}\left({\frac {H_{\rm h}}{R}}\right)^{-2},    \label{kh_condi}
\end{equation}
is satisfied. For $\alpha=0.1$ and $H_{\rm h}/R=0.5$, $\beta_{{\rm h},r}\le \beta_{{\rm h},r}^{\rm cr}=1600$ is required to suppress the instability, which can be easily satisfied in a differentially rotating   disk \citep[][]{1991ApJ...376..214B}. It indicates that the cold clumps with a weak magnetic field are immune to the KH instability due to the radial motion of ambient hot gas, which implies the clumps can survive in the  radially inflowing hot gas without being dissolved.



\end{document}